\documentclass[preprint,prb,aps]{revtex4-1}

\usepackage{graphicx}
\usepackage{dcolumn}   
\usepackage{bm,braket}        
\usepackage{amssymb}   
\usepackage{amsmath}
\usepackage{url,color}
\usepackage{circuitikz}
\usepackage{qcircuit}
\usepackage{CJKutf8}
\usepackage{layout}
\usepackage{booktabs}
\usepackage{float}
\usepackage{titlesec}
\usepackage{multirow}

\def\be{\begin{equation}}
\def\ee{\end{equation}}
\def\ba{\begin{eqnarray}}
\def\ea{\end{eqnarray}}
\def\sdg{Schr\"odinger~}

\begin{document}
\title{Quantum Entanglement and Chocolates }
\begin{CJK*}{UTF8}{gbsn}
\author{Biao Wu (吴飙)}
\email{wubiao@pku.edu.cn}
\affiliation{International Center for Quantum Materials, School of Physics, Peking University,  Beijing 100871, China}
\affiliation{Wilczek Quantum Center, Shanghai Institute for Advanced Study, 
Shanghai 201315, China}
\affiliation{Hefei National Laboratory, Hefei 230088, China}
\date{\today}

\begin{abstract}
Two statistical ensembles of chocolates are constructed to mimic two quantum entangled states, 
the spin singlet state and the GHZ state. Despite great efforts
to achieve the closest possible  resemblance  in terms of 
probability distribution and  correlation,  subtle differences remain.  
The differences reveal and illustrate a fundamental  characteristic that
distinguishes  quantum entanglement  from classical correlation. 
\end{abstract}
\maketitle
\end{CJK*}

\section{Introduction with \sdg}
\label{sec:intro}

Although quantum mechanics was fully established by 1926, 
it was not until 1935 that physicists began to notice quantum entanglement. 
Inspired by the famous EPR thought experiment \cite{EPR}, 
Schr\"odinger first defined and discussed quantum entanglement\cite{schrod} in a 
paper published in 1935.  Schr\"odinger began his paper with the following,
\begin{quote}
When two systems, of which we know the states by their respective representatives, enter into 
temporary physical interaction due to known forces between them, and when after a time 
of mutual influence the systems separate again, then they can no longer be described in the 
same way as before, viz. by endowing each of them with a representative of its own. 
I would not call that one but rather the characteristic trait of quantum mechanics, 
the one that enforces its entire departure from classical lines of thought. 
By the interaction the two representatives (or $\psi$-functions) have become entangled. 
\end{quote}
The ``representative" or $\psi$-function used by Schr\"odinger refers to pure quantum state 
or wave function  in modern terminology.  Here Schr\"odinger pointed out a fundamental feature
of quantum entanglement: any of the subsystems in an entangled state does not have a definite
pure quantum state. He asserted that this feature marks the complete departure of quantum mechanics
from classical mechanics. \\

Schr\"odinger was very insightful and his  view is profound. For a classical system, 
to know the system as a whole,  we have to know each of its components. 
For example, for a system with $n$ classical  particles, only after we know the position ${\bm x}_i$ and 
momentum ${\bm p}_i$ of each particle, we can claim that  
we know exactly what the state of the whole system is. This is consistent with our everyday experience. 
If you say that you know everything about a box of chocolates, everyone expects  you to know the type, 
the shape, and other properties of each piece of chocolate in the box. 
But for quantum entangled states, the situation  becomes completely different: 
it is possible know the system as a whole without knowing its parts. \\

Any quantum entangled state is sufficient to demonstrate this quite striking feature. I choose 
the spin singlet state,  
\be
\label{eq:singlet}
\ket{\psi}=\frac{1}{\sqrt{2}}(\ket{\uparrow\downarrow}-\ket{\downarrow\uparrow})\,.
\ee
Here the spin-up state $\ket{\uparrow}$ and spin-down state $\ket{\downarrow}$ are two eigenstates 
of $\hat{\sigma}_z$.  Each spin lives in a two-dimensional Hilbert space ${\mathcal H}_2$;
the two spins live in the composite Hilbert space ${\mathcal H}={\mathcal H}_2\otimes {\mathcal H}_2$.
$\ket{\psi}$ is a  vector in Hilbert space ${\mathcal H}$ that gives a complete 
description of the state of the two spins.  
They are in a superposition of two states: in the first state spin 1 is up and spin 2 is down; in the second state 
spin 1 is down and spin 2 is up. But what are the states of spin 1 and spin 2, respectively? 
They no longer have their ``respective representatives" according to \sdg or pure quantum states 
in modern terminology. According to quantum mechanics\cite{chuang}, the state of spin 1 is found 
by tracing out spin 2 in $\ket{\psi}$, giving us  the following mixed state, 
\be
\label{eq:dmx}
\rho=\frac12 \ket{\uparrow}\bra{\uparrow}+\frac12 \ket{\downarrow}\bra{\downarrow}\,.
\ee
Mathematically the state $\rho$ is not a vector in Hilbert space ${\mathcal H}_2$. 
Physically, it differs from any pure single-spin 
state. One may argue that $\rho$ is physically the same as  
\be
\ket{\varphi}=\frac{1}{\sqrt{2}}(\ket{\uparrow}+\ket{\downarrow})\,,
\ee
because when the spin is measured along the $z$ direction,  both $\rho$ and $\ket{\varphi}$ 
have a 50\% chance of getting spin up or down. However, $z$ is a special direction; 
the results are different for other directions. For example,  
for the $x$ direction, $\rho$ still has a 50\% chance of getting either positive or negative  outcome; 
in contrast, $\ket{\varphi}$ has a 100\% chance of finding the spin in the 
positive $x$ direction. 
If we could prepare two pieces of chocolate to similar quantum entangled states, 
then each  chocolate would no longer have a definite type or shape.\\

\sdg at the time also noticed another feature of quantum entanglement,  correlation at distance. 
He continued in his 1935 paper\cite{schrod},
\begin{quote}
Of either system, taken separately, all previous knowledge may be entirely lost, 
leaving us but one privilege: to restrict the experiments to one only of the two systems. 
After re-establishing one representative by observation, the other one can be inferred simultaneously.
\end{quote}
It shows that \sdg  not only noticed the correlation between two entangled systems 
but also clearly indicated with ``simultaneously" that 
it is independent of distance between the two subsystems. It is a correlation at distance. 
This is clearly the case for  the singlet state (\ref{eq:singlet}), which contains no information
about the distance between the two spins.   
However, besides the above short passage, Schr\"odinger had little discussion of this correlation   
in the paper.  It is now impossible to know exactly why Schr\"odinger 
did not pay much attention to this property. 
One possible reason is that Schr\"odinger did not find it as striking since correlation
at distance is quite common in everyday life. 
Mike is on a business trip. After checking into his hotel, Mike opens his suitcase and finds 
that he has only the left-handed (right-handed) glove.
Then no matter how far away from home he is, he knows instantly that 
the glove left at home must be  right-handed (left-handed).  
\sdg may have thought that the correlation in quantum entanglement was similar 
to the correlation between the gloves. 
We now know that there is a very subtle but fundamental 
difference between them, which was   revealed 
by Bell in 1964 with an inequality \cite{bell}. 
I will discuss this distinction in detail in this work. 

Beside the spin singlet state  (\ref{eq:singlet}), I will discuss another well known entangled state, 
the GHZ state,\cite{GHZ,Mermin}
\be
\label{eq:ghz}
\ket{\Phi}=\frac{1}{\sqrt{2}}\big(\ket{\uparrow\uparrow\uparrow}-
\ket{\downarrow\downarrow\downarrow}\big)\,.
\ee
I will construct two statistical ensembles of chocolates to resemble these two quantum entangled states 
as much as possible in terms of probability distribution and correlation.   
Despite great efforts, there always remain some crucial differences, which mark the fundamental 
departure of quantum mechanics from classical mechanics as \sdg asserted.

\section{The spin singlet and the classical ensemble}
\label{sec:singlet}
The spin singlet state  has a very important property - rotational symmetry. 
Eq. (\ref{eq:singlet}) is its expression  in terms of the eigenstates of $\hat{\sigma}_z$. 
The expression remains unchanged if it is instead  written in terms of 
the eigenstates of the  spin operator $\vec{n}\cdot\hat{\sigma}$ along an arbitrary direction $\vec{n}$. 
That is Eq. (\ref{eq:singlet}) can be re-written as
\be
\ket{\psi}=\frac{1}{\sqrt{2}}(\ket{n_+n_-}-\ket{n_-n_+})\,,
\ee
where $\ket{n_+}$ and $\ket{n_-}$ are two eigenstates of  $\vec{n}\cdot\hat{\sigma}$ and they are given by
\be
\ket{n_+}=
\begin{pmatrix}
e^{-i\varphi/2}\cos\frac{\theta}{2}\\e^{i\varphi/2}\sin\frac{\theta}{2}
\end{pmatrix}\,,~~~~\ket{n_-}=\begin{pmatrix}
-e^{-i\varphi/2}\sin\frac{\theta}{2}\\e^{i\varphi/2}\cos\frac{\theta}{2}
\end{pmatrix}\,.
\ee
This  means that measuring  the spin along any direction $\vec{n}$, 
if the measurement result for spin 1 is  positive, then spin 2 must point in the negative direction of $\vec{n}$. 
Conversely, if the measurement result for spin 1 is negative, then spin 2 must point 
in the positive direction of $\vec{n}$.  Similarly, the corresponding mixed state (\ref{eq:dmx}) becomes
\be
\label{eq:dmx2}
\rho=\frac12 \ket{n_+}\bra{n_+}+\frac12 \ket{n_-}\bra{n_-}\,.
\ee
This means that measuring either spin 1 or spin 2 along an arbitrary direction $\vec{n}$, 
the result is 50\% of positive and 50\% of negative. 

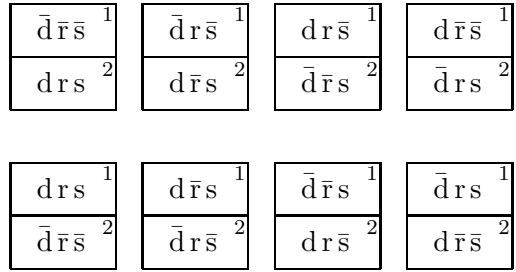
\begin{figure}[h]
\begin{picture}(100,90)(80,0)
  \linethickness{0.15mm}
  \newsavebox{\hezi}
\savebox{\hezi}
 (90,20)[bl]{
  \multiput(0,0)(40,0){2}%
    {\line(0,1){40}}
 \multiput(0,0)(0,20){3}%
    {\line(1,0){40}}
    \put(34.7,13.0){\scriptsize{2}}
    \put(34.7,33.5){\scriptsize{1}}
    }
   \multiput(30,0)(50,0){4}%
    {\usebox{\hezi}}  
    \multiput(30,60)(50,0){4}%
    {\usebox{\hezi}}  
    \put(40,27.0){d\,r\,s}
    \put(40,7.0){$\bar{\rm d}$\,$\bar{\rm r}$\,$\bar{\rm s}$}
    \put(90,27.0){d\,$\bar{\rm r}$\,s}
    \put(90,7.0){$\bar{\rm d}$\,r\,$\bar{\rm s}$}
    \put(140,27.0){$\bar{\rm d}$\,$\bar{\rm r}$\,s}
    \put(140,7.0){d\,r\,$\bar{\rm s}$}
    \put(190,27.0){$\bar{\rm d}$\,r\,s}
    \put(190,7.0){d\,$\bar{\rm r}$\,$\bar{\rm s}$}
     \put(40,87.0){$\bar{\rm d}$\,$\bar{\rm r}$\,$\bar{\rm s}$}
    \put(40,67.0){d\,r\,s}
    \put(190,87.0){d\,$\bar{\rm r}$\,$\bar{\rm s}$}
    \put(190,67.0){$\bar{\rm d}$\,r\,s}
     \put(90,87.0){$\bar{\rm d}$\,r\,$\bar{\rm s}$}
    \put(90,67.0){d\,$\bar{\rm r}$\,s}
     \put(140,87.0){d\,r\,$\bar{\rm s}$}
    \put(140,67.0){$\bar{\rm d}$\,$\bar{\rm r}$\,s}
\end{picture}
\caption{Eight ways of  boxing chocolates in the singlet ensemble. d is for dark, $\bar{\rm d}$  for 
non-dark (or milk); r is for round, $\bar{\rm r}$ for non-round; s is for Swiss, $\bar{\rm s}$ for non-Swiss. 
Note that the chocolate in compartment 1 of the first box may be milk, sea-shell-shaped, 
and made in Denmark.}
\label{singletc}
\end{figure}

I try to construct  an ensemble of  chocolates that has similar statistical properties as 
the spin singlet.  It consists of $2N$ chocolates, of which $N$ chocolates are dark, 
$N$ chocolates are round,  and $N$ chocolates are Swiss. Note that some of the 
chocolates may be milk, square-shaped, and made in Belgium. 
In addition, there are $N$ identical boxes, each with two compartments. 
We put the chocolates into the boxes in pairs. The boxing rule is:
the two chocolates in the same box can NOT be both dark, round, and Swiss. 
According to this rule, there are eight ways of pairing chocolates in a box as shown in
Figure \ref{singletc}. A pair of chocolates is put in a box randomly in one of the eight ways. 
Let us compare the statistical properties of this chocolate ensemble 
with those of the spin singlet. \\

{\it Single-body probability}~-~For the spin singlet (\ref{eq:singlet}), 
measuring  spin 1 along any direction $\vec{n}$, there are two possible outcomes: 
a 50\% chance of being positive and a 50\% chance of being negative. 
In the chocolate ensemble, there is a 50\% chance that the chocolate in compartment 1 
is dark and a 50\% chance that it is not; a 50\% chance that it is round 
and a 50\% chance that it is not; and a 50\% chance that it is Swiss and a 50\% chance that it is not.
It is obvious that the spin singlet and the  chocolate ensemble are identical 
in terms of the single-body probability. 
Comparison of  spin 2 and chocolates in compartment 2 has similar results. 

{\it Correlation at distance}~-~As the chocolates in the ensemble are boxed 
according to a given rule as illustrated in Figure \ref{singletc}, 
there is correlation between the chocolates in the same box. 
If the chocolate in compartment 1 is dark, then the chocolate in compartment 2 of the same 
box must not be dark. Conversely, if the chocolate in compartment 1 is not dark, 
then the other chocolate  must be dark. 
Although the two compartments are next to each other, the correlation between the chocolates is 
distance-independent: imagine that the two chocolates in the box are 
put into two identical small boxes, covered, and separated. No matter how far apart, 
when you find  the chocolate in one of the small boxes is not dark, 
the chocolate  in the other small box far away from you must be dark. \\

Obviously, the conclusion is the same for the other two properties of chocolate, 
shape and production place. 
If the chocolate in compartment1 is triangular and made in Switzerland, 
then the chocolate in compartment 2 must be round and produced in a country 
other than Switzerland. \\

Similar correlation exists between spin 1 and spin 2 in the  singlet state (\ref{eq:singlet}) 
as already discussed above: 
if the measurement finds spin 1 up, then spin 2 must be down; 
if the measurement finds spin 1 down, then spin 2 must be up. \\

The above analysis seems to indicate that there is no difference between the spin singlet 
and the corresponding chocolate ensemble not only in terms of single-body probability   but also 
from the point of view of  correlation. This is not true. 
The difference does exist, but it is more subtle and must be carefully analyzed. 
One has to resort to Bell's inequality that I will discuss below.\\

{\it Bell's Inequality}~-~Discovered by Bell in 1964\cite{bell}, this inequality can demonstrate 
a subtle but profound difference between the spin singlet and the chocolate ensemble. 
Bell's inequality is concerned with the correlation between the different properties of two chocolates, 
such as the probability that one chocolate  is dark and the other in the same box is round. 
For simplicity and convenience, we let dark be property $A$, round be property $B$, and Swiss be property $C$. 
At the same time, we introduce three correlation probabilities: 
$p(A,B)$, $p(B,C)$, and $p(A,C)$.  $p(A,B)$ is the probability that the chocolate in 
compartment 1 is dark and the chocolate in compartment 2 in the same box is round; 
$p(B,C)$ is the probability of finding a box
where the chocolate in compartment 1 is round and the chocolate in compartment 2 is Swiss; 
$p(A,C)$ is the probability  that the chocolate in  compartment 1 is dark 
and the chocolate in the other compartment of the same box is Swiss.  
Bell shows that these three probabilities must satisfy the following inequality,
\be
\label{eq:bell1}
p(A,B)+p(B,C)\ge p(A,C)\,.
\ee
The proof of this inequality is not difficult and will be given below.\\

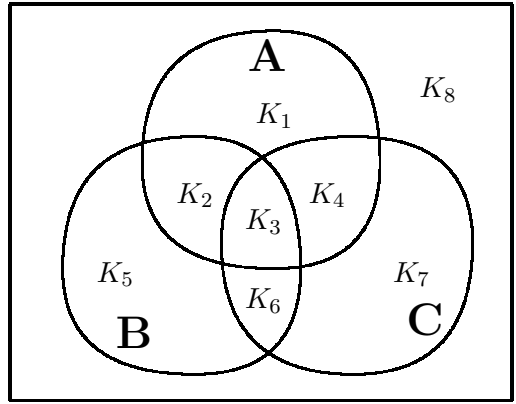
\begin{figure}[h]
\begin{picture}(200,150)(10,0)
  \linethickness{0.35mm}
  \multiput(0,0)(190,0){2}%
    {\line(0,1){150}}
 \multiput(0,0)(0,150){2}%
    {\line(1,0){190}}
     \linethickness{0.25mm}
  \qbezier(110,50)(110,100)(70,100)
  \qbezier(70,100)(20,100)(20,50)
  \qbezier(20,50)(20,10)(70,10)
  \qbezier(70,10)(110,10)(110,50)  
   \put(40,20){\Large{\bf B}}
  %
  \qbezier(175,60)(175,100)(130,100)
  \qbezier(130,100)(80,100)(80,60)
  \qbezier(80,60)(80,10)(130,10)
  \qbezier(130,10)(175,10)(175,60)  
 \put(150,25){\Large{\bf C}}
   \qbezier(140,90)(140,140)(100,140)
  \qbezier(100,140)(50,140)(50,90)
  \qbezier(50,90)(50,50)(100,50)
  \qbezier(100,50)(140,50)(140,90)  
  \put(90,125){\Large{\bf A}}
  \put(93,105){$K_1$}
  \put(63,75){$K_2$}
  \put(89,65){$K_3$}
  \put(113,75){$K_4$}
  \put(33,45){$K_5$}
  \put(89,35){$K_6$}
  \put(145,45){$K_7$}
  \put(155,115){$K_8$}
  \end{picture}
\caption{The large square  represents all the  elements in a set. 
The upper circle is the subset of all elements with property A;
the lower left circle is the subset of elements with property B; the lower right circle 
is the subset of elements with property C. For chocolates, the large square represents 
the entire chocolate ensemble: A for dark, B for round, and C for Swiss. 
For spins, the large square represents all the measurement results: A denotes positive measurement results
along the $\vec{n}_1$ direction, B denotes positive along $\vec{n}_2$, and C denotes positive 
along  $\vec{n}_3$.}
\label{belld}
\end{figure}

We notice that if the chocolates in compartment 2 are round, 
then according to the boxing rule the chocolate in compartment 1 in the same box 
must not be round. So, if we define $\tilde{p}(A,\neg B)$ as the probability 
that the chocolate in compartment 1 is dark but not round, 
then it is clear that $\tilde{p}(A,\neg B)$ is the same as $p(A,B)$, i.e.,  
$\tilde{p}(A,\neg B)=p(A,B)$.  Similarly, we have $\tilde{p}(B,\neg C)=p(B,C)$ 
and $\tilde{p}(A,\neg C)=p(A,C)$.  Thus the inequality (\ref{eq:bell1}) is equivalent to
\be
\label{eq:bell2}
\tilde{p}(A,\neg B)+\tilde{p}(B,\neg C)\ge \tilde{p}(A,\neg C)\,.
\ee
Note that $p(A,B)$ is the probability of correlation between two chocolates, 
while $\tilde{p}(A,\neg B)$ is the probability of a single chocolate. These two probabilities 
are equal because of the correlation  between two chocolates in the same box. If the chocolates
were boxed without any rule, we would not have $\tilde{p}(A,\neg B)=p(A,B)$. \\

We prove the inequality (\ref{eq:bell2}) with the help of Figure \ref{belld}~\cite{wu}. 
The large square in Figure \ref{belld} represents all the $N$ chocolates in compartment 1.
This large square is divided into eight parts by the three properties of chocolates: 
$K_1$, $K_2$, $K_3$, $K_4$, $K_5$, $K_6$, $K_7$, and $K_8$. Each part
is a subset of chocolates that have certain shared properties. 
For example, $K_1$ includes all dark chocolates that are not round and  not made in Switzerland, 
$K_3$ is all dark Swiss chocolates that are round, $K_8$ contains all the chocolates 
that are not dark, not round, and non-Swiss. 
According to Figure \ref{belld}, we have
\begin{align}
\tilde{p}(A,\neg B)&=\frac{K_1+K_4}{2N}\,,~~~\\
\tilde{p}(B,\neg C)&=\frac{K_2+K_5}{2N}\,,~~~\\
\tilde{p}(A,\neg C)&=\frac{K_1+K_2}{2N}\,.
\end{align}
Therefore, we obtain 
\be
\tilde{p}(A,\neg B)+\tilde{p}(B,\neg C)=\frac{K_1+K_2+K_4+K_5}{2N}\ge \tilde{p}(A,\neg C)\,.
\ee 
This proves the inequality (\ref{eq:bell2}), and consequently  Bell's inequality (\ref{eq:bell1}). 
To illustrate, we use  the chocolates in the eight boxes of Figure \ref{singletc} as an example. 
For the particular case,  we have
\be
p(A,B)=\frac14\,,~~p(B,C)=\frac14\,,~~p(A,C)=\frac{1}{4}\,. 
\ee
They clearly satisfy Bell's inequality (\ref{eq:bell1}).\\

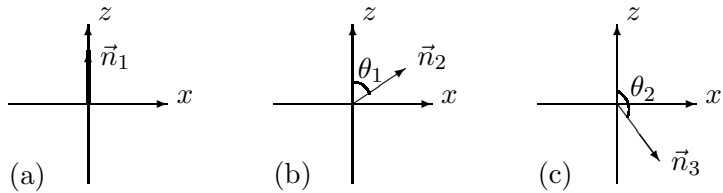
\begin{figure}[h]
\begin{picture}(160,60)(50,0)
  \newsavebox{\xzzb}
\savebox{\xzzb}(90,20)[bl]{
\linethickness{0.25mm}
  \put(-30,0){\vector(1,0){60}}
\put(33.5,0.5){$x$}
\put(0,-30){\vector(0,1){60}}
\put(3.5,32){$z$}
}
\multiput(30,0)(100,0){3}%
    {\usebox{\xzzb}}  
     \linethickness{0.5mm}
    \put(30,30){\vector(0,1){20}}
    \put(34.5,45){$\vec{n}_1$}
    \put(0,0){(a)}
    \linethickness{0.5mm}
    \put(130,30){\vector(3,2){20}}
    \put(154.5,45){$\vec{n}_2$}
    \put(132,40){$\theta_1$}
    \linethickness{0.2mm}
    \qbezier(130,38)(133.5,39)(136.5,34)
    \put(100,0){(b)}
    \linethickness{0.5mm}
    \put(230,30){\vector(3,-4){16}}
    \linethickness{0.2mm}
    \qbezier(230,35)(236,31.5)(234,25)
    \put(250.5,5){$\vec{n}_3$}
    \put(235,33){$\theta_2$}
    \put(200,0){(c)}    
\end{picture}
\caption{Three directions: $\vec{n}_1$, $\vec{n}_2$, and $\vec{n}_3$. }
\label{n1n2n3}
\end{figure}

Surprisingly,  the quantum correlation in the spin singlet (\ref{eq:singlet}) may violate 
Bell's inequality (\ref{eq:bell1}). To show this, we consider three directions in the $xz$-plane:
$\vec{n}_1$, $\vec{n}_2$, and $\vec{n}_3$ as shown in Fig. \ref{n1n2n3},  
and measure the spin along these three directions. 
For comparison with chocolates, we label the positive result 
of  measuring spin along the direction of $\vec{n}_1$ as property A.
Similarly, the positive result of the spin along the  direction of $\vec{n}_2$ is labeled property B; 
the positive result of the spin along the $\vec{n}_3$  direction is labeled as property C. \\

We measure  spin 1 along the $\vec{n}_1$ direction, and  spin 2 along the $\vec{n}_2$ direction. 
(Note that the  results are independent of the order of measurements.)
The probability that both the measurement results are positive is
\be
p_q(A,B)=\frac12\sin^2\frac{\theta_1}{2}\, .
\ee
Similarly, we measure spin 1 along the $\vec{n}_2$ direction and  spin 2 along the $\vec{n}_3$ direction.
The probability that both results are positive is 
\be
p_q(B,C)=\frac12\sin^2\frac{\theta_2-\theta_1}{2}\,.
\ee
We measure spin 1 along the $\vec{n}_1$ direction and spin 2 along the $\vec{n}_3$ direction.
The probability that both results are positive is
\be
p_q(A,C)=\frac12\sin^2\frac{\theta_2}{2}\,.
\ee
Let us choose $\theta_1=\pi/3$ and $\theta_2=2\pi/3$. In this case,  we have
\be
p_q(A,B)=\frac18\,,~~~p_q(B,C)=\frac18\,,~~~p_q(A,C)=\frac38\,.
\ee
Obviously, 
\be
p_q(A,B)+p_q(B,C)< p_q(A,C)\,.
\ee
This shows that the correlation probability between two spins  in the spin singlet 
can violate Bell's inequality (\ref{eq:bell1})! 
This possibility of violating Bell's inequality by the spin singlet is its subtle and fundamental
difference from the chocolate ensemble.\\

Note that different values of $\theta_1$ and $\theta_2$ correspond to  boxing  chocolates 
in the ensemble differently. We proved earlier that Bell's inequality is satisfied 
no matter how the chocolates are boxed, as long as the rule of boxing is satisfied. 
Why is Bell's inequality violated by the singlet for some values of $\theta_1$ and $\theta_2$. 
 Let us now analyze the violation more closely. 

 Bell's inequality is proved with the help of Fig. \ref{belld}. We will see that 
 this figure is not well defined for spin. To be specific, the total set is defined as 
 all the possible measurement results for spin 1 and it
is represented by the large square in Fig. \ref{belld}. In this figure, circle A is the subset of 
all the positive measurement results for spin 1 along the $\vec{n}_1$ direction; 
circle B is the subset of 
all the positive measurement results for spin 1 along the $\vec{n}_2$ direction; 
circle C is the subset of 
all the positive measurement results for spin 1 along the $\vec{n}_3$ direction. 

A smaller subset, which consists of  two parts, $K_1$ and $K_4$, contains 
all  the measurement results for spin 1 that are positive along the $\vec{n}_1$ direction and simultaneously negative along the $\vec{n}_2$  direction. This subset is in a sense not well defined 
because in general one can not measure a single spin along two different directions at the same time.
Fortunately, for the special case of the singlet state, this subset is well defined.  
As spin 1 is in a singlet state, the negative results along the $\vec{n}_2$ 
direction can be inferred from the positive results for the measurement on spin 2 along the $\vec{n}_2$ 
direction. In other words, one can measure spin 1 along $\vec{n}_1$ 
direction and simultaneously measure spin 2 along $\vec{n}_2$ direction. If both outcomes are positive, 
they belong to the subset made of $K_1$ and $K_4$. 

When we examine the smallest subsets, for example, $K_1$, we are in real trouble. 
$K_1$ is supposed to contain certain outcomes when spin 1 is measured simultaneously along three different directions: $\vec{n}_1$, $\vec{n}_2$, and $\vec{n}_3$.  This is impossible, and even the 
singlet state can not come to rescue. The last-ditch rescue effort is to allow 
 the three measurements be done at different times; still $K_1$ is ill-defined 
 because the results depend on the order of the measurements.  
If  the measurement were done first along $\vec{n}_1$, then $\vec{n}_2$, and finally $\vec{n}_3$, then the size of $K_1$
would be proportional to $\sin^2\frac{\theta_1}{2}\cos^2\frac{\theta_2-\theta_1}{2}$; if the order of $\vec{n}_2$ 
and $\vec{n}_3$ were switched, then $K_1$ proportional to $\sin^2\frac{\theta_2}{2}\cos^2\frac{\theta_2-\theta_1}{2}$. Two different orders two different results, so the subset $K_1$ is not well  defined. 

So, all the small subsets $K_j, j=1,2,\cdots,8$ in Fig. \ref{belld} are not well defined for spins. 
The fundamental reason is that the three operators, 
 $\vec{n}_1\cdot\hat{\sigma}$, $\vec{n}_2\cdot\hat{\sigma}$, and $\vec{n}_3\cdot\hat{\sigma}$, 
 do not commute with each other; as a result, the overall measurement outcome depends on the time order 
 of the three measurements. 
 For chocolates, all these subsets $K_j, j=1,2,\cdots,8$ are well defined. 
This is because the different properties  of chocolates 
can be  possessed and inspected (or measured) simultaneously;  
 once the chocolates are boxed, the sizes of these subsets are determined.

\section{The GHZ state and the classical ensemble}
\label{sec:ghz}
The GHZ state (\ref{eq:ghz}) is another quantum state often used to illustrate 
quantum entanglement. It is an entangled state with three spins. 
We are only concerned here with the measurement  of each spin along the $x$, $y$, and $z$ directions.  
It is clear that any one of the spins in the GHZ state 
no longer has a definite pure quantum state, instead it is in a mixed state (\ref{eq:dmx}). 
As a result, if we measure any of the three spins along the $x$, $y$, or $z$ direction, 
we have a  50\% chance of getting positive results and a  50\% chance of getting negative results. \\

There are correlations among the spins. When  measuring any spin in the GHZ state along the $z$ direction, 
if the result is up, then the results for the other two spins are also up; if  the result is down, the other two are down. 
 Either of the two outcomes has a 50\% chance. The correlations along the $x$ and $y$ directions 
 are not as straightforward. They are implied in the  following equations, 
\begin{align}
\label{eq:xyy1}
\hat{\sigma}_x^{(1)}\hat{\sigma}_y^{(2)}\hat{\sigma}_y^{(3)}|\Phi\rangle &=|\Phi\rangle\,, \\
\label{eq:xyy2}
\hat{\sigma}_y^{(1)}\hat{\sigma}_x^{(2)}\hat{\sigma}_y^{(3)}|\Phi\rangle &=|\Phi\rangle\,,\\
\label{eq:xyy3}
\hat{\sigma}_y^{(1)}\hat{\sigma}_y^{(2)}\hat{\sigma}_x^{(3)}|\Phi\rangle &=|\Phi\rangle\,,
\end{align}
which can be verified by direct computation. 
 These three equations show that,   
 if we measure one of the three spins along the $x$ direction and the other two spins in the $y$ direction, 
 each of the three results can be positive or negative, but it must be positive when they are multiplied together. 
 An equivalent statement is that the following two measurement outcomes are impossible: 
 (1) one result is negative and the other two are positive; (2) all three results are negative.\\

\begin{figure}[h]
\begin{picture}(90,110)(80,0)
  \linethickness{0.15mm}
  \newsavebox{\gezi}
\savebox{\gezi}
 (90,20)[bl]{
  \multiput(0,0)(30,0){4}%
    {\line(0,1){20}}
 \multiput(0,0)(0,20){2}%
    {\line(1,0){90}}
    \put(24.7,14.0){\scriptsize{1}}
    \put(54.7,14.0){\scriptsize{2}}
    \put(84.7,14.0){\scriptsize{3}}
    }
 \multiput(30,0)(0,30){4}%
    {\usebox{\gezi}}   
 \multiput(170,0)(0,30){4}%
    {\usebox{\gezi}}    
  \put(37,7.0){d\,r}
  \put(13,7.0){(7)}
    \put(67.7,7.0){d\,r}
    \put(97.7,7.0){d\,r}
    \put(177,7.0){d\,$\bar{\rm r}$}
    \put(153,7.0){(8)}
    \put(207.7,7.0){d\,$\bar{\rm r}$}
    \put(237.7,7.0){d\,$\bar{\rm r}$}
    \put(37,37.0){d\,$\bar{\rm r}$}
    \put(13,37.0){(5)}
    \put(67.7,37.0){$\bar{\rm d}$\,r}
    \put(97.7,37.0){$\bar{\rm d}$\,r}
    \put(37,67.0){$\bar{\rm d}$\,r}
     \put(13,67.0){(3)}
        \put(13,97.0){(1)}
    \put(67.7,67.0){d\,$\bar{\rm r}$}
    \put(97.7,67.0){$\bar{\rm d}$\,r}
    \put(37,97.0){$\bar{\rm d}$\,r}
    \put(67.7,97.0){$\bar{\rm d}$\,r}
    \put(97.7,97.0){d\,$\bar{\rm r}$}
    \put(177,37.0){d\,r}
     \put(153,37.0){(6)}
    \put(207.7,37.0){$\bar{\rm d}$\,$\bar{\rm r}$}
    \put(237.7,37.0){$\bar{\rm d}$\,$\bar{\rm r}$}
    \put(177,67.0){$\bar{\rm d}$\,$\bar{\rm r}$}
      \put(153,67.0){(4)}
         \put(153,97.0){(2)}
    \put(207.7,67.0){d\,r}
    \put(237.7,67.0){$\bar{\rm d}$\,$\bar{\rm r}$}
      \put(177,97.0){$\bar{\rm d}$\,$\bar{\rm r}$}
    \put(207.7,97.0){$\bar{\rm d}$\,$\bar{\rm r}$}
    \put(237.7,97.0){d\,r}
\end{picture}
\caption{Eight ways to box the chocolates in the  GHZ ensemble. d is for dark, 
$\bar{\rm d}$ for non-dark; r is for round, $\bar{\rm s}$ for other shapes. 
For example, for the box in (1), the chocolates in both compartments 1 and 2 are round and milk;
the  chocolate in compartment 3 is dark but not round. 
In four of the eight boxes, all the  chocolates are Swiss; in the other four, all the
chocolates are not Swiss.}
\label{ghzc}
\end{figure}
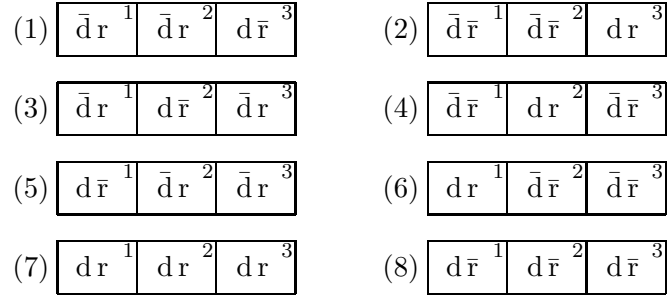

I construct an ensemble of chocolates to share these features of the GHZ state: 
the spin measurements along the $x$, $y$, and $z$ directions correspond to 
the three properties of chocolates, type, shape, and production place, respectively.  
There are $24N$ chocolates in the ensemble, of which $12N$ are dark, $12N$  round, and  $12N$ Swiss. 
These chocolates are placed  in $8N$ boxes with three in each box. 
Every box has three compartments, numbered 1, 2, and 3 from left to right. 
The $8N$ boxes are divided equally into eight groups, and each group is boxed in a different way. 
The eight ways of boxing are shown in Figure \ref{ghzc}.  
Below we compare the properties of this chocolate ensemble with those of the spin GHZ state.\\

{\it Single-body probability}~-~There are 24 chocolates in Figure \ref{ghzc}, 
of which $12$ chocolates are dark, $12$ chocolates are not dark. 
Therefore, in the whole chocolate ensemble, 
there are $12N$ chocolates that are dark and $12N$ chocolates that are not dark.
This means that there is a 50\% chance that the chocolate is dark and a 50\% chance that it is not black. 
With respect to shape and production place, 
we have similar results: there is a 50\% chance that the chocolate is round and a 50\% chance that it is not; 
there is a 50\% chance that the chocolate is Swiss and a 50\% chance that it is not Swiss.  
This is similar to the measurement of a single spin in the GHZ state (\ref{eq:ghz}).  
For measurements of any single spin in the GHZ state along the $x$,   $y$, or $z$ direction, 
the results are: 50\% chance of being positive and 50\% chance of being negative. 
Thus, in terms of the single-body probability, 
the  GHZ state and the corresponding chocolate ensemble are identical. \\

{\it Correlation at distance}~-~
Since the chocolates are packed in boxes according to the eight designed ways in Figure \ref{ghzc}, 
there is correlation among the three chocolates in the same box.
Clearly,  this correlation is independent of distance:
if the chocolates are separated and placed on three distant planets, 
the correlation among the three chocolates in the same box does not change. 
With respect to production place, the correlation 
is obvious: if any of the chocolates in a box is Swiss, the other two are also Swiss; 
if any of the chocolates in a box is not Swiss, the other two are not Swiss, either.  
This is similar to the measurement of spin in the GHZ state along $z$ direction.  

The correlations with respect to type and shape are more subtle. 
To demonstrate them, we define a correlation variable
\be
C_{xyy}=c_x^{(1)}c_y^{(2)}c_y^{(3)}\, .
\ee
If the chocolate in compartment 1 is dark, then $c_x^{(1)}=1$; if the chocolate in compartment 1 is not dark, 
then $c_x^{(1)}=-1$. If the chocolates in compartments 2 and 3 are round, then $c_y^{(2)}=c_y^{(3)}=1$;
if they are not round, then $c_y^{(2)}=c_y^{(3)}=-1$. For the first way of boxing in Figure \ref{ghzc}, 
we have $c_x^{(1)}=-1$, $c_y^{(2)}=1$, and $c_y^{(3)}=-1$, which lead to $C_{xyy}=1$. 
One can check that, for the other seven ways of boxing, we have the same result, $C_{xyy}=1$. 
This is consistent with the correlation in quantum entanglement embodied in Eq.(\ref{eq:xyy1}).\\

Similarly, we can define
\be
C_{yxy}=c_y^{(1)}c_x^{(2)}c_y^{(3)}\,,~~~C_{yyx}=c_y^{(1)}c_y^{(2)}c_x^{(3)}\,. 
\ee
A case-by-case check confirms that we have $C_{yxy}=C_{yyx}=1$ for all the boxings in Figure \ref{ghzc}. 
This is consistent with the quantum correlations (\ref{eq:xyy2},\ref{eq:xyy3}) in the GHZ state. \\

Although the chocolate ensemble is similar to the GHZ state in all aspects that we have examined, 
there is a crucial difference.  To show the difference, we define a new correlation variable
\be
C_{xxx}=c_x^{(1)}c_x^{(2)}c_x^{(3)}\, .
\ee
For the second way of boxing in Figure \ref{ghzc}, we have $c_x^{(1)}=-1$, $c_x^{(2)}=-1$,  
and $c_x^{(3)}=1$, which lead to  $C_{xxx}=1$. For the other seven cases, we similarly have $C_{xxx}=1$. 
This means that for this chocolate ensemble, $C_{xxx}=1$. In contrast, for the GHZ state, 
we can show by straightforward computation that
\be
\label{eq:xxx}
\hat{\sigma}_x^{(1)}\hat{\sigma}_x^{(2)}\hat{\sigma}_x^{(3)}|\Phi\rangle =-|\Phi\rangle\, .
\ee
This is the opposite of $C_{xxx}=1$. 
This suggests that the chocolate ensemble constructed based on Fig. \ref{ghzc}, while agreeing 
with the GHZ state in many aspects, is not be completely consistent with the GHZ state.
In fact, it can be rigorously proved that there does not exist a classical ensemble
that is identical to the GHZ state in terms of its probability distribution and correlation. 
If we assume that such an ensemble exists, to be consistant with the quantum correlations
indicated in Eqs. (\ref{eq:xyy1},\ref{eq:xyy2},\ref{eq:xyy3}), 
 this classical ensemble must satisfy the following three conditions:
\begin{align}
C_{xyy}&=c_x^{(1)}c_y^{(2)}c_y^{(3)}=1\,,\\
C_{yxy}&=c_y^{(1)}c_x^{(2)}c_y^{(3)}=1\,,\\
C_{yyx}&=c_y^{(1)}c_y^{(2)}c_x^{(3)}=1\,. 
\end{align}
Here $c_x$ and $c_y$ can be any two distinct properties of the system in the ensemble. 
Utilizing the equalities, 
\be
(c_y^{(1)})^2=(c_y^{(2)})^2=(c_y^{(3)})^2=1\,,
\ee
we can prove that
\begin{align}
&C_{xyy}C_{yxy}C_{yyx}\nonumber\\
=&c_x^{(1)}c_y^{(2)}c_y^{(3)}c_y^{(1)}c_x^{(2)}c_y^{(3)}c_y^{(1)}c_y^{(2)}c_x^{(3)}\nonumber\\
=&c_x^{(1)}c_x^{(2)}c_x^{(3)}=C_{xxx}\,.
\end{align}
Thus $C_{xxx}=1$. The chocolate ensemble we constructed does indeed satisfy this equation. In sharp contrast, 
the GHZ state does not. The reason is that 
the spin operators $\hat{\sigma}_x$ and $\hat{\sigma}_y$ do not commute, and 
Eq. (\ref{eq:xxx}) cannot be derived from Eqs. (\ref{eq:xyy1},\ref{eq:xyy2},\ref{eq:xyy3}).

\section{Summary}
\label{sec:diss}
I have constructed two chocolate ensembles, with the aim to to make them share as many features 
 as possible with their corresponding quantum entangled states.
 But I can not make the chocolate ensembles completely identical to the quantum entangled states.  
 This is a matter of principle. This is similar to that 
 we can never build a machine of perpetual motion because it would violate the first or second law of thermodynamics. 
 For any quantum entangled state, we can construct a classical ensemble whose one-body probability is identical to that of the entangled state, but  its many-body probability correlation cannot be exactly the same as that of the entangled state. 
 The fundamental reason is that different physical properties of a quantum system are related to 
different operators, which do not commute in general. As a result, the quantum correlation between these properties 
are fundamentally different from the classical correlation, which is showcased by the violation of Bell's inequality.


Although Schr\"odinger pointed out the most essential features of quantum entanglement, he did not 
propose a quantitative description of entanglement. 
In 1957 Everett described entanglement quantitatively in terms of von Neumann entropy 
in his doctoral thesis \cite{DeWitt,many}.

\begin{acknowledgments}
This work was supported by National Natural Science Foundation of China (Grants No.  11921005, No. 12475011, and No.  92365202, ), and Shanghai Municipal Science and 
Technology Major Project (Grant No.2019SHZDZX01). The author has no conflicts to disclose. 
\end{acknowledgments}

%
%

\begin{thebibliography}{99}
\expandafter\ifx\csname natexlab\endcsname\relax\def\natexlab#1{#1}\fi
\expandafter\ifx\csname bibnamefont\endcsname\relax
  \def\bibnamefont#1{#1}\fi
\expandafter\ifx\csname bibfnamefont\endcsname\relax
  \def\bibfnamefont#1{#1}\fi
\expandafter\ifx\csname citenamefont\endcsname\relax
  \def\citenamefont#1{#1}\fi
\expandafter\ifx\csname url\endcsname\relax
  \def\url#1{\texttt{#1}}\fi
\expandafter\ifx\csname urlprefix\endcsname\relax\def\urlprefix{URL }\fi
\providecommand{\bibinfo}[2]{#2}
\providecommand{\eprint}[2][]{\url{#2}}

\bibitem{EPR} A. Einstein, B. Podolsky, and N. Rosen, {\it Can Quantum-Mechanical Description of Physical Reality Be Considered Complete?}, 
Physical  Review, {\bf 47}, 777 (1935).

\bibitem{schrod}E. Schr\"odinger, {\it Discussion of Probability Relations between Separated Systems}, Proceedings of the Cambridge Philosophical Society, {\bf 31} 555 (1935).

\bibitem{chuang} M. A. Nielsen and I. L. Chuang, {\it Quantum Computation and Quantum Information}
 (Cambridge University Press, Cambridge, UK, 2000).

\bibitem{bell} J.S. Bell, {\it On the Einstein-Podolsky-Rosen paradox}, Physics {\bf 1}, 195 (1964). 

 \bibitem{GHZ}D. M. Greenberger, M. A. Horne, A. Shimony, and A. Zeilinger, {\it Bell's theorem without inequalities}, 
 American Journal of Physics {\bf 58}, 1131 (1990). 
 
 \bibitem{Mermin} N. David Mermin, {\it What's Wrong with these Elements of Reality?}, 
 Physics Today {\bf 38}, 38 (2008). 

\bibitem{wu} Biao Wu, {\it Quantum Mechanics: A Concise Introduction} 
(Springer, Singapore, 2023).

\bibitem{DeWitt}B.S. DeWitt, N. Graham (eds.), The Many-Worlds Interpretation of Quantum Mechanics (Princeton Uni- versity Press, Princeton, 1973)
\bibitem{many} Biao Wu, {\it Everett's theory of the universal wave function}, 
The European Physical Journal H {\bf 46}, 7 (2021). 
 

 




 %
\end{thebibliography}

\end{document}